\documentclass{article}

\usepackage[preprint]{musicscore}

\usepackage[utf8]{inputenc} 
\usepackage[T1]{fontenc}    
\usepackage{hyperref}       
\usepackage{url}            
\usepackage{booktabs}       
\usepackage{amsfonts}       
\usepackage{nicefrac}       
\usepackage{microtype}      
\usepackage{xcolor}         
\usepackage{enumitem}
\usepackage{graphicx}
\usepackage{musicography}   
\usepackage{subcaption}

\usepackage{amsmath}

\title{MusicScore: A Dataset for Music Score Modeling and Generation}

\author{%
  Yuheng Lin\thanks{\url{https://rozenthegoat.github.io/}} \\
  Department of Information Engineering \\
  The Chinese University of Hong Kong \\
  \texttt{yuhenglin@link.cuhk.edu.hk} \\
  \And
  Zheqi Dai \\
  Department of Electronic Engineering \\
  The Chinese University of Hong Kong \\
  \texttt{dzq1387848806@gmail.com} \\
  \And
  Qiuqiang Kong\thanks{Corresponding author.} \\
  Department of Electronic Engineering \\
  The Chinese University of Hong Kong \\
  \texttt{qqkong@ee.cuhk.edu.hk} \\
}

\begin{document}

\maketitle

\begin{abstract}

Music scores are written representations of music and contain rich information about musical components. The visual information on music scores includes notes, rests, staff lines, clefs, dynamics, and articulations. This visual information in music scores contains more semantic information than audio and symbolic representations of music. Previous music score datasets have limited sizes and are mainly designed for optical music recognition (OMR). There is a lack of research on creating a large-scale benchmark dataset for music modeling and generation. In this work, we propose MusicScore, a large-scale music score dataset collected and processed from the International Music Score Library Project (IMSLP). MusicScore consists of image-text pairs, where the image is a page of a music score and the text is the metadata of the music. The metadata of MusicScore is extracted from the general information section of the IMSLP pages. The metadata includes rich information about the composer, instrument, piece style, and genre of the music pieces. MusicScore is curated into small, medium, and large scales of 400, 14k, and 200k image-text pairs with varying diversity, respectively. We build a score generation system based on a UNet diffusion model to generate visually readable music scores conditioned on text descriptions to benchmark the MusicScore dataset for music score generation. MusicScore is released to the public at \url{https://huggingface.co/datasets/ZheqiDAI/MusicScore}.

\end{abstract}

\section{Introduction}
\label{sec:intro}

Music scores \cite{jones2017musicowl, cook2013beyond, dannenberg2006music} are written representations of music and contain rich information about music. The visual information in music scores includes notes, rests, staff lines, clefs, dynamics, and articulations. This visual information has advantages over audio \cite{briot2017deep, dieleman2018challenge} and symbolic representations \cite{ji2023survey, zeng2021musicbert} of music in many music modeling and generation tasks, as it contains rich semantic information. Composers create music by writing music scores, making them the most direct modality for representing music. However, there is a lack of datasets and benchmarks for music score generation. In this work, we propose MusicScore, a large-scale music score dataset and a diffusion model-based music score generation system to benchmark music score generation.

Music datasets are fundamental for music modeling and generation. Previous music datasets can be categorized into \textit{audio datasets} and \textit{symbolic music datasets}. Audio datasets contain the raw audio of music and have been widely used for music generation. Examples of audio datasets include FMA \cite{defferrard2016fma}, Music4All \cite{santana2020music4all}, and Disco10M \cite{lanzendorfer2024disco}. Audio datasets have also been applied to music information retrieval tasks, such as music transcription \cite{benetos2018automatic, hawthorne2017onsets, kong2021high}, music tagging \cite{choi2017convolutional}, chord estimation \cite{weiss2020local}, and source separation \cite{lu2024music}. Audio datasets have the advantage of containing rich audio information, such as timbre and dynamics, which are difficult to represent in symbolic music datasets. However, audio datasets lack abstract musical information, such as key signatures, time signatures, notes, and rests, which are important for music creation and analysis.

Symbolic music datasets encode music information into a structured and symbolic format, such as MIDI \cite{yang2017midinet, chou2021midibert}, MusicXML \cite{good2001musicxml}, ABC notation \cite{geerlings2020interacting}, Humdrum \cite{huron2002music}, LilyPond \cite{nienhuys2003lilypond}, and Octopus \cite{liu2020octopus}. Symbolic music datasets describe music in terms of abstract symbols that represent musical elements such as notes and chords. Symbolic music datasets include MAESTRO \cite{hawthorne2018enabling}, Lakh MIDI Dataset \cite{manilow2019cutting}, GiantMIDI-Piano \cite{kong2020giantmidi}, JSB Chorales \cite{peracha2021js}, KernScores \cite{sapp2005online}, MuseData \cite{hewlett199727}, and others. Symbolic music datasets are typically more compact in size compared to audio datasets and can be easily edited and manipulated. Symbolic music datasets high-level musical information that is useful for tasks such as music theory analysis and algorithmic music generation. However, symbolic music datasets are usually limited in scale and quality compared to audio datasets. Many symbolic music datasets crawled or transcribed from the internet may contain noise.

Previous music score datasets are mainly designed for optical music recognition (OMR), such as DeepScores \cite{tuggener2018deepscores, tuggener2021deepscoresv2}, DoReMi \cite{guo2023doremi}, and Universal Music Symbol Collection \cite{pacha2017towards}. However, those OMR datasets only contain labels for recognition purposes. Many datasets such as DeepScores \cite{tuggener2018deepscores, tuggener2021deepscoresv2} are rendered from MIDI or MusicXML datasets and lack variety. Additionally, these OMR datasets contain a limited number of composers and music pieces, which restricts their applications for music modeling and generation. Furthermore, these OMR datasets lack metadata information about the scores, such as composers, instruments, era, and text descriptions. There is a lack of large-scale music score image-text pair datasets for music modeling and generation. 

In this work, we focus on creating MusicScore, a large-scale music score dataset containing image-text pairs for music modeling and generation. The contributions of this dataset are as follows. First, we are the first to propose the \textit{music score generation} dataset and benchmark. We propose that music score generation is a novel way of generating music from the vision modality. Second, we collect MusicScore by downloading, processing, and cleaning music scores and their corresponding metadata from the International Music Score Library Project (IMSLP) \cite{imslp}, the largest public music collection in the world. Third, we curate MusicScore into small, medium, and large scales of 400, 14k, and 200k image-text pairs with various diversity, respectively. Fourth, we process MusicScore to contain rich metadata information collected from the general information section of each piece on IMSLP, including but not limited to genre, instrumentation, piece style, etc. Fifth, we build a text-driven latent diffusion system to generate high-quality and playable sheet music images aligned with textual descriptions to benchmark music score generation systems. We release both the MusicScore dataset and the processing code to the public to facilitate research on music score modeling and generation.

This paper is organized as follows. Section 2 introduces related works on music generation. Section 3 introduces the MusicScore dataset. Section 4 introduces the music score generation benchmark system. Section 5 discusses the limitations of MusicScore. Section 6 concludes this work.

\section{Related works of music generation}
\label{sec:related}

Previously, there have been two major research directions to address the music generation problem, including audio music generation and symbolic music generation.

\paragraph{Audio music generation} 

Audio music generation systems generate music waveforms and audio representations. These systems can be trained end-to-end to produce audio signals without the need for intermediate symbolic music representations. WaveNet \cite{oord2016wavenet} is an representative music generation system modeled by dilated convolutional neural networks (CNNs). SampleRNN \cite{mehri2016samplernn} applies an improved hierarchical recurrent neural network (RNN) to model waveforms. Recently, large language models have been applied to music generation, such as MusicLM \cite{agostinelli2023musiclm}. Diffusion models are types of probabilistic generative model that incrementally transforms noise into music through a series of learned denoising steps. Diffusion model-based music generation systems include MusicLDM \cite{chen2024musicldm}, AudioLDM \cite{liu2023audioldm, liu2024audioldm}, MeLoDy \cite{lam2024efficient}, and music consistency models \cite{fei2024music}. MusicRL \cite{cideron2024musicrl} aligns music generation with human preferences. The advantage of audio music generation systems lies in their universality. However, using audio music generation systems to generate long music waveforms may result in missing semantic information and is not easy to edit and manipulate.

\paragraph{Symbolic music generation} 

Symbolic music generation \cite{ji2023survey} generate music in symbolic formats, such as MIDI and ABC notations. The music symbols include the instrument pitch, onset, velocity, and duration information of music. In symbolic music generation, the music symbols are transformed into a series of tokens and input to a language model. Music Transformer \cite{huang2018music} is the first work to apply Transformer \cite{vaswani2017attention} for symbolic piano music generation. Transformers-based methods have been widely used for symbolic music generation in \cite{muhamed2021symbolic, shih2022theme}. Diffusion models are applied for symbolic music generation in \cite{mittal2021symbolic}. MuseGAN \cite{dong2018musegan} is a multiple-track music generation system that can generate four bars of music. The generated symbolic music has the advantage of being easy to edit, transpose, and manipulate. However, symbolic music data generation systems lack expressive information of music and require high-quality symbolic music data for training. The generated music symbols also lack performance articulations, which are important for musicians to play.

\section{MusicScore dataset}
\label{sec:dataset}

Different from audio datasets and symbolic music datasets, music scores provide human-readable visual representations. The visual information in scores is important for musicians to analyze and perform music. To fill the gap between the visual modality and music modality for music generation, we collect MusicScore, a large-scale image-text pair dataset for music score modeling and generation. The MusicScore dataset consists of high-quality A4-sized classical music score images. These images are obtained by processing music score PDF files from IMSLP \cite{imslp}. The metadata for each music piece is obtained by collecting the corresponding general information section from the IMSLP page. MusicScore is carefully filtered and cleaned to ensure the quality of the metadata. We describe the collection and processing of MusicScore as follows.

\begin{figure}[t]
  \centering
  \includegraphics[width=1.0\textwidth]{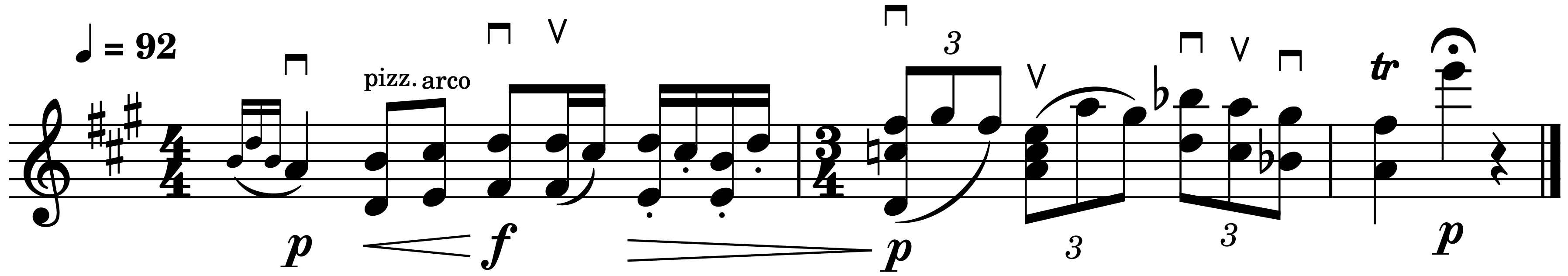}
  \caption{
  An example of an \textbf{A major violin score} demonstrates the fundamental elements of staff notation, including clefs, accidentals, dynamics and other performance techniques notation for bowed string instruments.
  }
  \label{fig:sheet-score}
\end{figure}

\subsection{Musical semantics}

The semantic music symbols on music scores convey important information about how a composer composes and what emotions the composer wishes to convey. The first level of semantic information is the organization of staffs on a music score. For example, a symphony may contain tens of lines of staffs to represent the staffs of multiple instruments. The instruments can be categorized into woodwind, brass, percussion, and string groups. A piano solo should contain both treble and bass clefs. Each score page may contain several blocks.

The secondary level of musical semantic information is within each line. The semantic symbols include the playing techniques and performance instructions that affect the timbre, speed, emotion, and sound of the music. Figure~\ref{fig:sheet-score} shows an example of a violin solo score. Starting from the leftmost part of each line on the staff, the first semantic symbol is the clef. The clef can be a treble clef \musSymbol{0.6em}{-0.0ex}{0.5em}{\symbol{71}}, a bass clef \musSymbol{0.6em}{1ex}{0.5em}{\symbol{73}}, or other clefs to locate pitches on a staff within a range. Following the clef symbol, there is a time signature $ 4 / 4 $ indicating there are 4 beats per measure. The key signature consists flat symbol \musFlat~or sharp symbol \musSharp~symbols, determining the key of a piece. Ornamentation includes notes that decorate another note, which cannot be easily transformed into symbolic representations. The directions of the beams indicate different parts of a music piece. In some polyphonic piano music, there can be multiple parts, such as a leading voice part and a bass voice part. The $p$ and $f$ symbols indicates the dynamics of music. The music semantic information may also contain texts, such as \textit{crecs.}, \textit{rit.} to indicate the dynamic and speed instructions of music, respectively. The texts \textit{pizz.} and \textit{arco} indicate the performance technique of string instruments. The slur symbol indicates that notes should be played without separation. The symbols \musSymbol{1em}{-0.0ex}{2em}{\symbol{22}}, \musSymbol{1em}{-0.0ex}{2em}{\symbol{23}}, $\cdot$, \musSymbol{1em}{-0.0ex}{2em}{\symbol{57}}, \musSymbol{1em}{-0.0ex}{2em}{\symbol{80}} are performance instructions.

\subsection{IMSLP}

The International Music Score Library Project (IMSLP) \cite{imslp}, commonly referred to as the Petrucci Music Library, is an esteemed subscription-based digital library that houses a vast collection of public-domain music scores and is renowned as the world's largest electronic sheet music archive. IMSLP serves as a valuable resource for musicians and scholars worldwide. By providing unrestricted access to an extensive repertoire of musical compositions, IMSLP promotes the dissemination and preservation of musical knowledge and heritage \cite{wiki_imslp}. 

\subsection{Dataset downloading}

\begin{figure}
    \centering
    \includegraphics[width=1.0\textwidth]{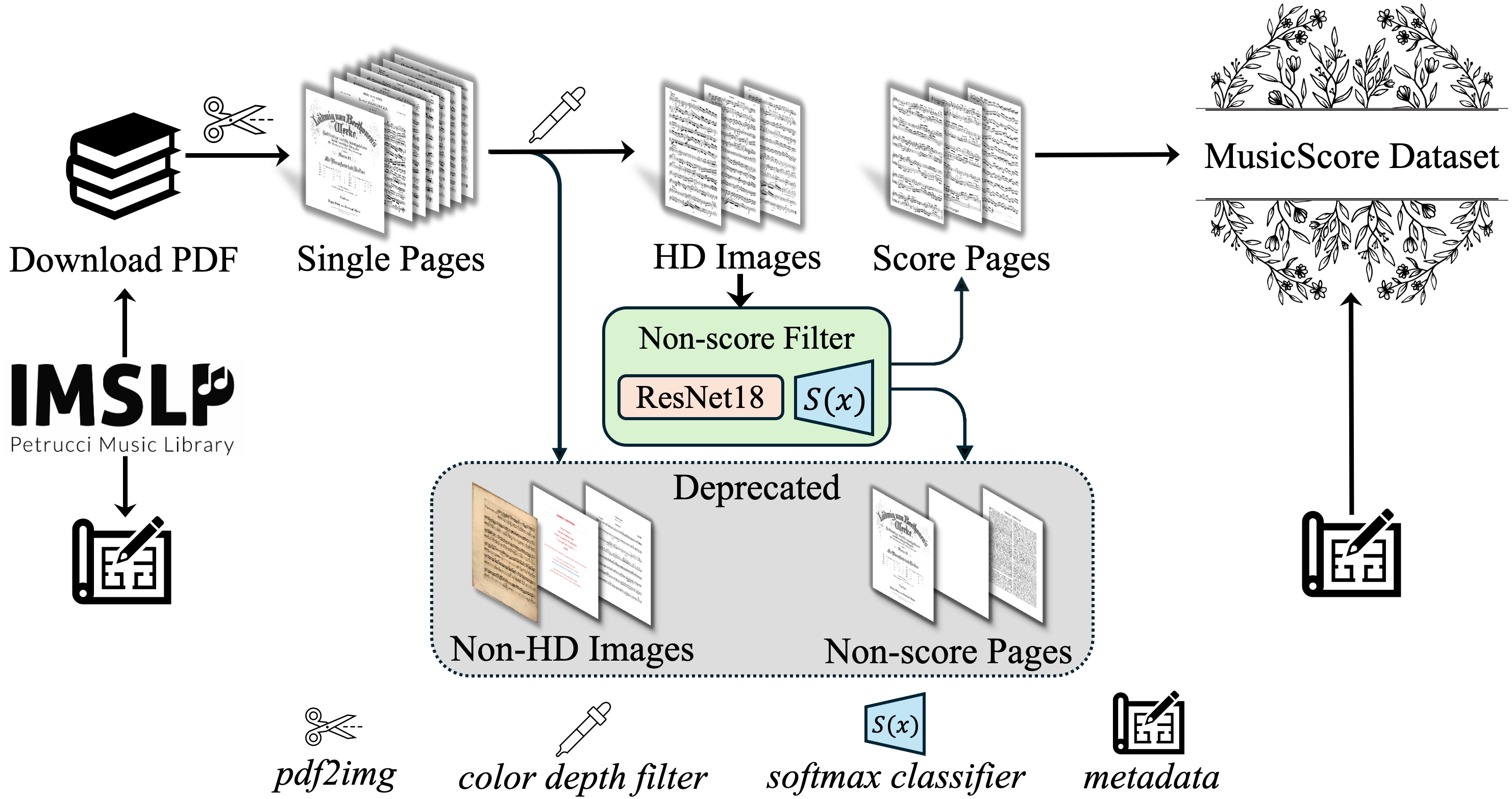}
    \caption{MusicScore dataset collecting and processing pipeline.}
    \label{fig:pipeline}
\end{figure}

Figure~\ref{fig:pipeline} shows the music collection and processing pipeline. We download music scores from the IMSLP website \cite{imslp} in PDF format. Each PDF file may contain multiple pages. In May 2024, we downloaded a total of 148,257 PDF music scores.

\subsection{Dataset processing}

\begin{figure}[t]
    \centering
    \begin{minipage}[b]{0.24\linewidth}
        \centering
        \includegraphics[width=\linewidth]{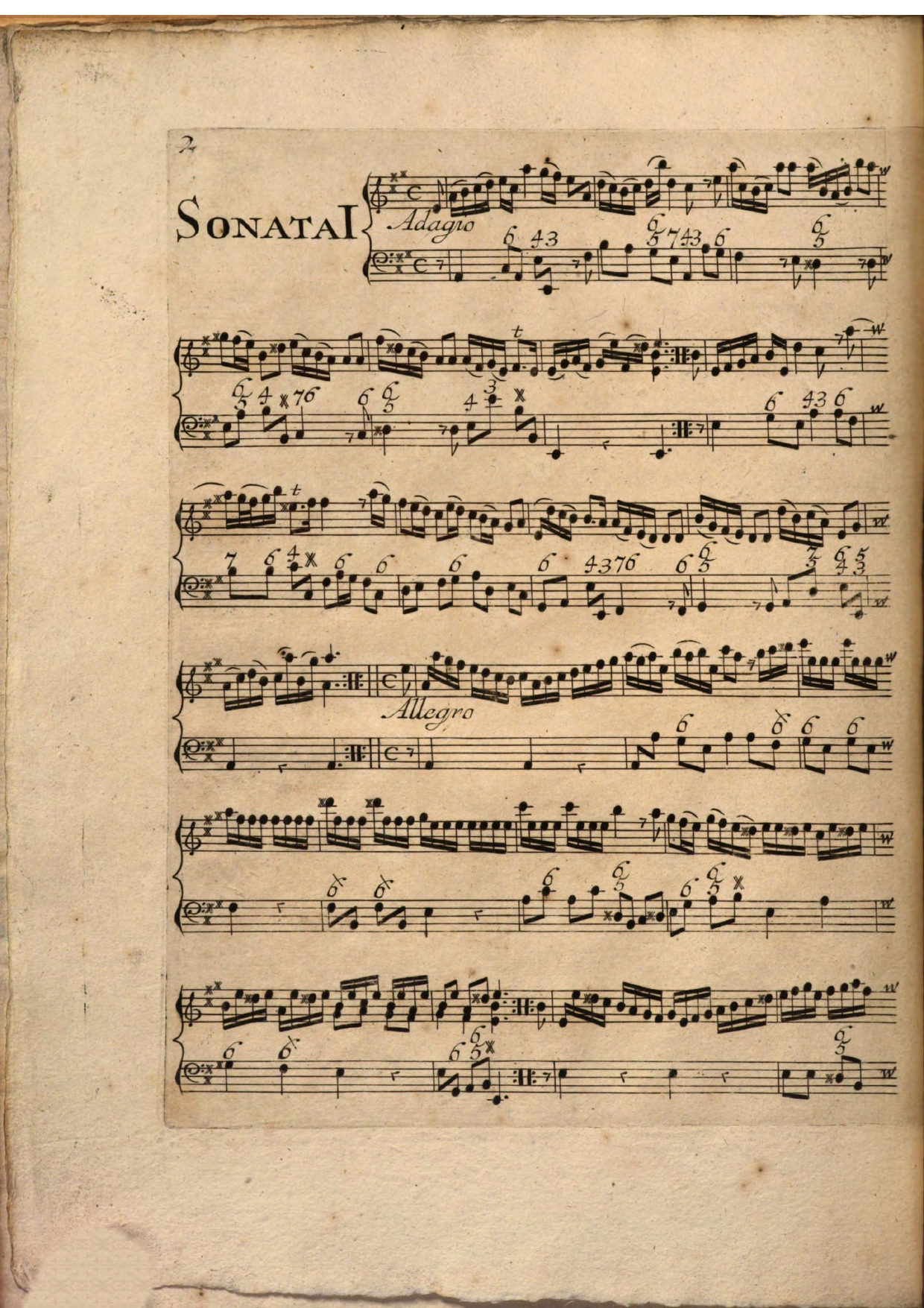}

    \end{minipage}
    \hfill
    \begin{minipage}[b]{0.24\linewidth}
        \centering
        \includegraphics[width=\linewidth]{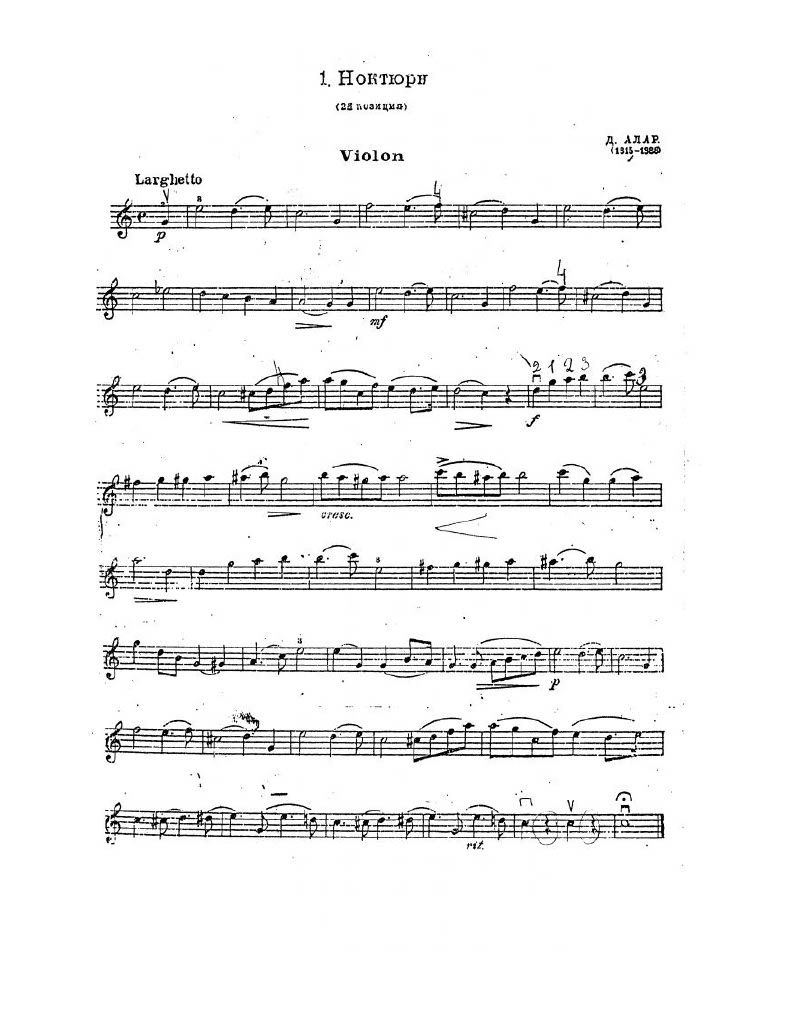}
    \end{minipage}
    \hfill
    \begin{minipage}[b]{0.24\linewidth}
        \centering
        \includegraphics[width=\linewidth]{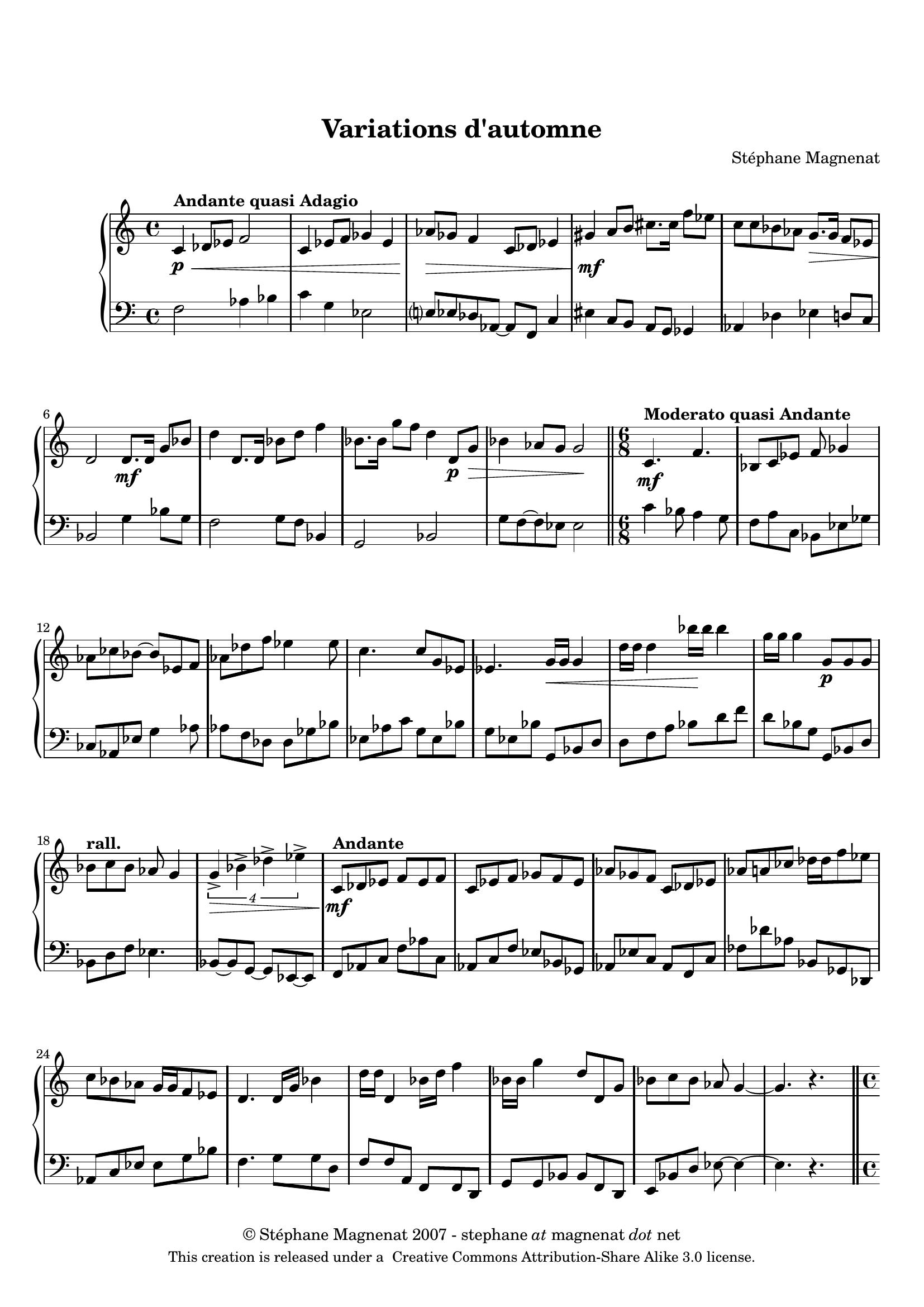}

    \end{minipage}
    \hfill
    \begin{minipage}[b]{0.24\linewidth}
        \centering
        \includegraphics[width=\linewidth]{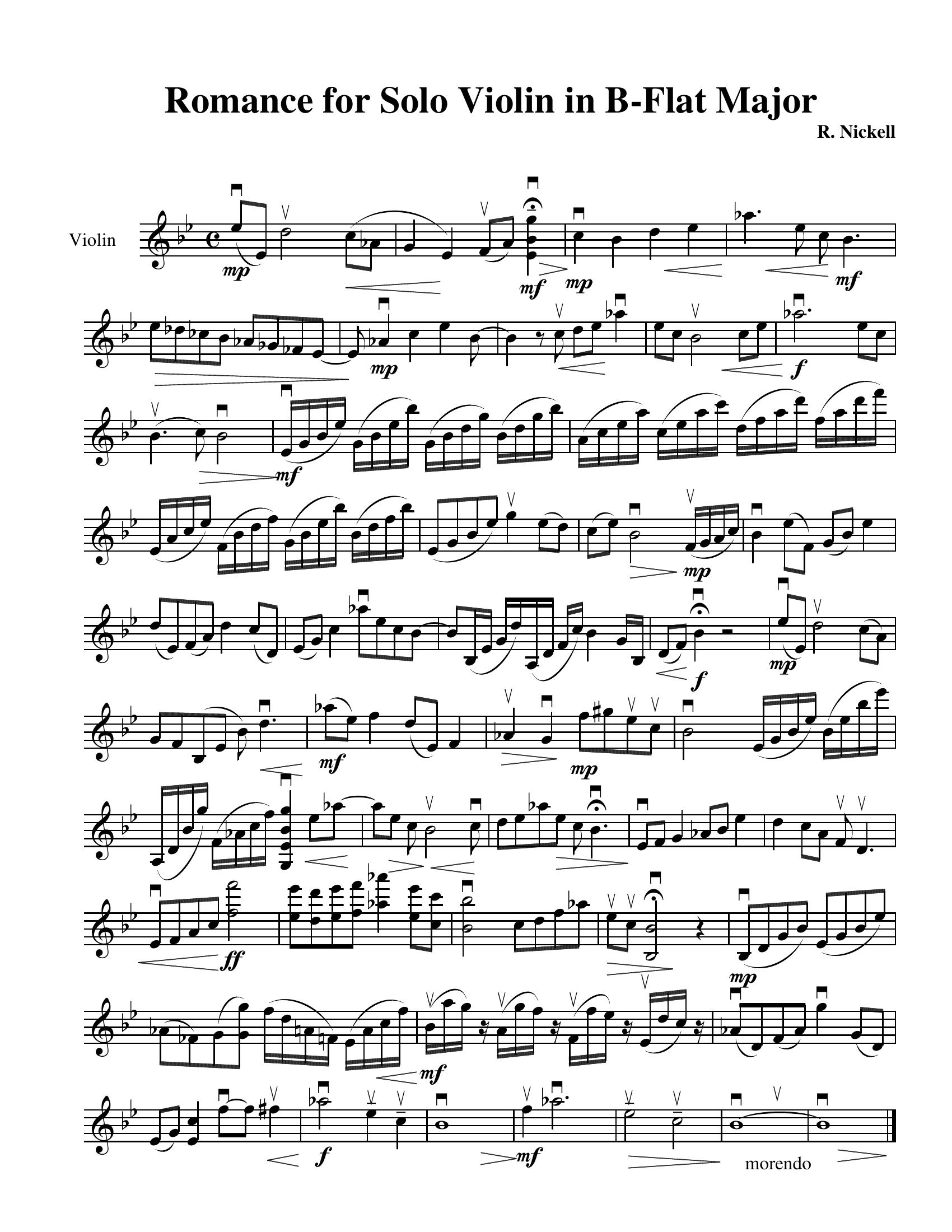}
    \end{minipage}
\caption{The first column: low-quality score with yellow pages. The second column: low-quality score with too much white spaces and noisy point. The third and fourth columns: high-quality scores.}
\label{fig:low_quality_images}
\end{figure}

Among those 148,257 PDF music scores, not all of them are of high-quality for music score generation. The music scores in IMSLP vary in quality. First, many handwritten scores are manuscripts written by composers who lived centuries ago. Second, many PDF files contain catalogs, textual descriptions, and other front pages that are not relevant to music scores. Third, there are also a number of scores that have been manually scanned and uploaded, containing page misalignment errors or unclear symbols due to low scanner resolution. Figure~\ref{fig:low_quality_images} shows examples of two low-quality and two high-quality scores, respectively. We designed the following criteria to clean the music scores.

\begin{itemize}[leftmargin=*]
    \item \textbf{High Quality}
    High-quality images are usually created by publishers using professional tools or software such as MuseScore \cite{musescore}, Sibelius \cite{sibelius}, and Finale \cite{finale}. The color of high-quality scores should be in grayscale with three RGB channels.

    \item \textbf{Full Page Score}
    In each page of a music score PDF, the proportion of musical content should occupy more than two-thirds of the page. A page should not contain too many texts irrelevant to the music content, such as those on the front cover page.
    
\end{itemize}

In order to create a high-quality single-page music score dataset that satisfies the above criteria, we applied three dataset processing stages as follows.

\subsubsection{Filter dataset by color depth}
\label{sec:binary}

We extract the color depth information of the PDF files. The color depth of 1-bit corresponds to black and white images, while the color depth of 8-bit or 16-bit corresponds to color images. Some existing works apply the binarization operation to convert color scores into black and white scores. However, the binarization operation often results in blurry or noisy black and white scores, with parts of notes and staves missing after binarization. To ensure the high quality of our dataset, we only retain the music scores with a color depth of 1-bit and remove the music scores with color depths larger than 1-bit. We implement this using the  \texttt{PyMuPDF}\footnote{\url{https://pypi.org/project/PyMuPDF/}} Python package. After filtering, there are 32,307 PDF files remained in the dataset.

\subsubsection{Filter dataset by removing non-score pages}

We partition all PDF files into separate pages by using the \textit{pdf2img}\footnote{\url{https://pypi.org/project/pdf2image/}} Python package. We name each page of a music score as \textit{scoreID\_pageIndex.jpg}, where \textit{scoreID} is the unique score ID and \textit{pageIndex} is the page index number, respectively. For example, the name \textit{IMSLP514102\_17.jpg} corresponds to the 17th page of the music piece \textit{IMSLP514102}. 

We train a classification model to classify score and non-score images. This model helps us exclude images that contain insufficient music content, such as cover pages and pages with textual descriptions, from the dataset. Through this step, we ensure that the selected single-page sheet music images only contain music contents. To train the classifier, we manually labelled 450 images, including 270 musical score images and 180 non-musical score images. Additionally, we manually label 50 images for testing purposes, including 30 musical score images and 20 non-musical score images. We resize the music scores to the resolution of $2048 \times 2048$, feeding them into a ResNet18 \cite{he2016resnet} with a cross-entropy loss to distinguish the score image from non-score images. Our model achieves an accuracy of 98\% on the test set. After removing the non-score pages, there are 200,480 images in JPG format remained. We refer to this dataset as MusicScore-200k. We then randomly sample 14,656 images from MusicScore-200k to create a medium-scale subset, MusicScore-14k, for the rapid development of music score generation systems.

\subsubsection{Manually labelled subset}

Training with large-scale datasets such as MusicScore-14k and -200k can be computationally intensive, not all users have access to sufficient computational resources. In order to lower the barrier for using MusicScore, we manually curated a small-scale dataset called MusicScore-400, consisting of 403 image-text pairs. The MusicScore-400 subset is not part of the MusicScore-200k. Instead, MusicScore-400 contains music scores partitioned from 19 compositions in history, composed by renowned musicians such as Bach, Beethoven, and Chopin. Among them, there are 3 complete piano pieces and 16 violin compositions. After partitioning the scores into separate images, the piano pieces consist of 215 pages, while the violin compositions consist of 188 pages. The metadata format of the MusicScore-400 dataset is consistent with MusicScore-14k and -200k. By providing this smaller dataset, we aim to make it more accessible for researchers with limited resources to explore and experiment with the music score image data, encouraging a wider range of engagement in tasks related to music score images and music generation.

\begin{figure}[t]
    \centering
    \begin{minipage}{0.7\textwidth}
        \centering
        \label{table:muisca-metadata}
        \begin{tabular}{|l|l|}
        \hline
        \textbf{Work Title} & Muisca - El Dorado \\
        \textbf{Alternative Title} & A cappella motet \\
        \textbf{Composer} & Steer, Michael Maxwell \\
        \textbf{I-Catalogue Number} & None [force assignment] \\
        \textbf{Key} & G minor \\
        \textbf{Year/Date of Composition} & 2018 \\
        \textbf{First Publication} & 2021 \\
        \textbf{Librettist} & Michael Maxwell Steer (b. 1946) \\
        \textbf{Language} & English \\
        \textbf{Average Duration} & 3'30" minutes \\
        \textbf{Composer Time Period} & Modern \\
        \textbf{Piece Style} & Modern \\
        \textbf{Instrumentation} & mixed chorus (SATB) a cappella \\
        \textbf{ID} & IMSLP690834 \\
        \hline
        \end{tabular}
    \end{minipage}%
    \begin{minipage}{0.3\textwidth}
        \centering
        \includegraphics[width=\linewidth]{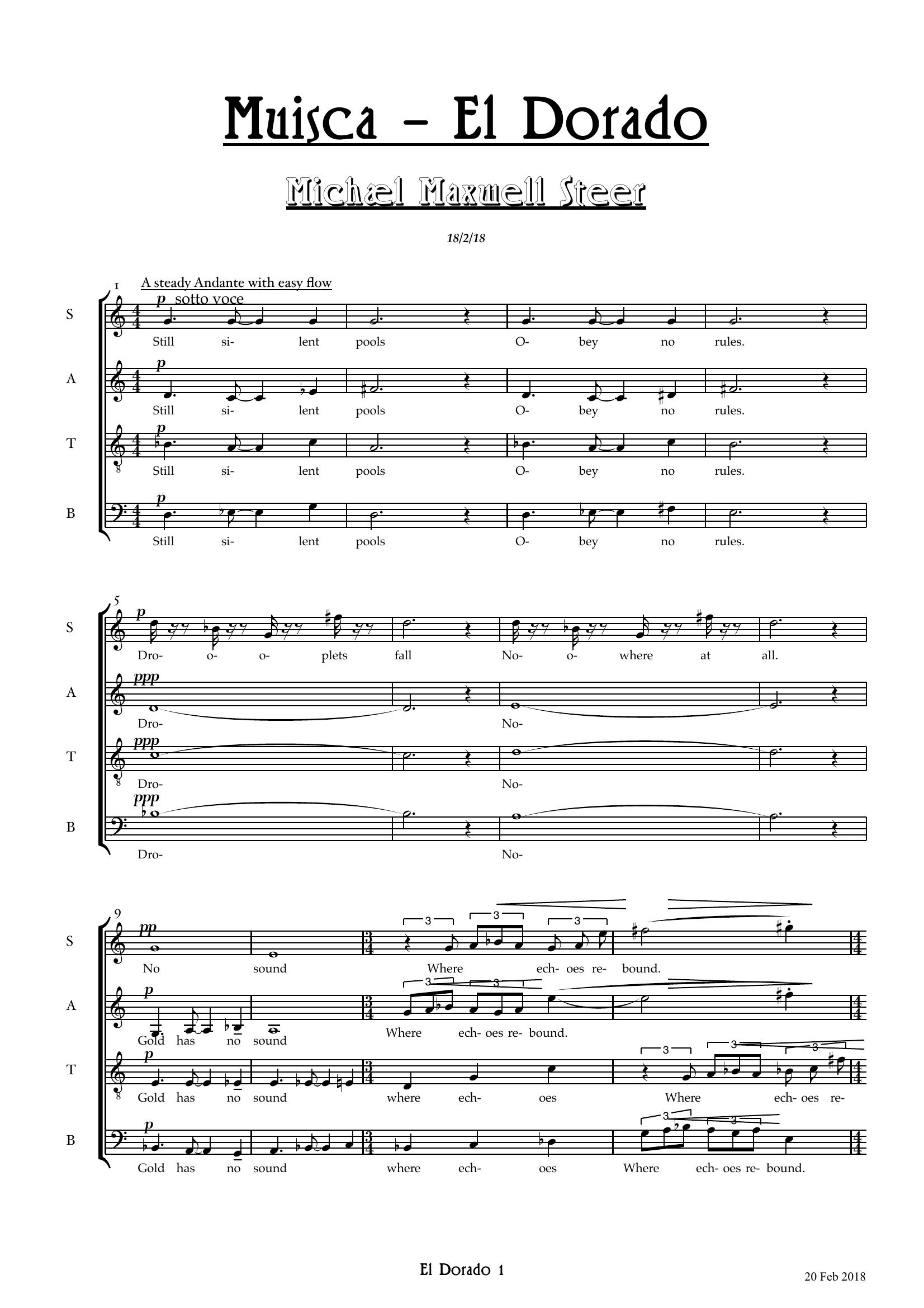} 
    \end{minipage}
    \caption{Left: metadata of ``Muisca - El Dorado'' by Michael Maxwell Steer. Right: The first page score.}
\label{fig:metadata}
\end{figure}

\subsection{Metadata processing}

We collect metadata from the general information section of each composition's IMSLP webpage. The metadata is stored in a JSON file using the data structure of a Python dictionary and is associated with the PDF files through name IDs. The metadata contains rich information such as the work title, composer, key, piece style, instrumentation, etc. Figure~\ref{fig:metadata} shows an example of metadata and the corresponding cover page of a score.

\subsection{Data statistics}
\label{sec:statistics}

\begin{figure}[t]
    \centering
    \includegraphics[width=1\linewidth]{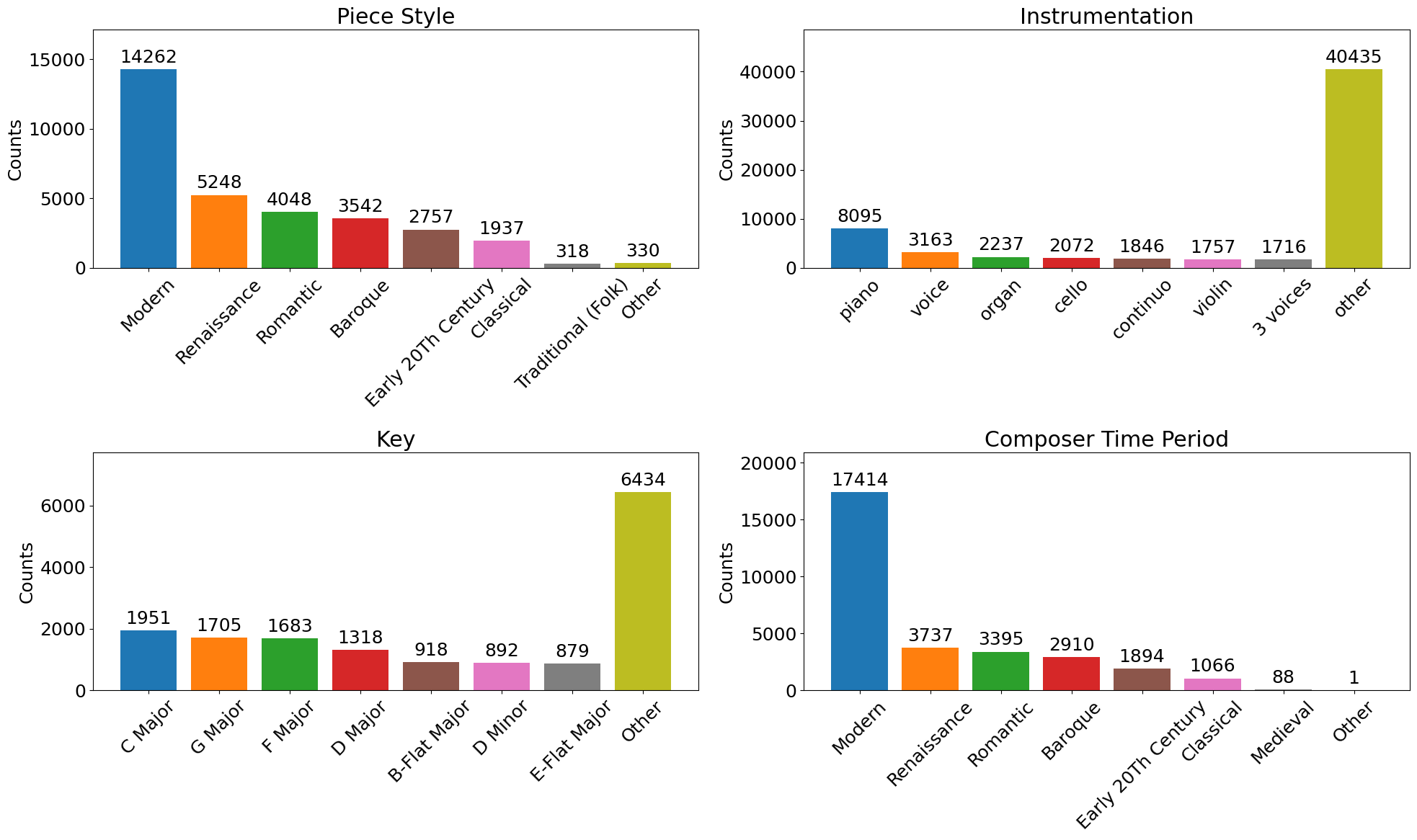}
    \caption{The data statistics of music scores in the MusicScore-200k dataset.}
    \label{fig:enter-label}
\end{figure}

We conducted a statistical analysis of MusicScore based on the metadata of music pieces. All statistics are calculated from the 32,307 PDF files. Figure~\ref{fig:enter-label} shows the statistics of music styles, instruments, keys, and composer time periods of the music scores. The top-left figure shows that the majority of piece styles are Modern, including 14,262 PDF files.
The top-right figure shows that MusicScore contains mostly piano scores with 8,095 PDF files. Since there can be multiple instruments in one music score, the number of other instruments is 40,435, larger than the total number of 32,307 PDF files. The bottom-left figure shows the top seven keys of the music scores. The top three keys are C major, G major, and F major. The bottom-right figure shows the composer period. There are 17,414 Modern composers in the dataset, followed by 3,737 Renaissance composers and 3,395 Romantic composers. The MusicScore dataset does not have negative impacts on society and can be freely used by the public under CC BY 4.0 standard.

\section{Experiments}
\label{sec:exp}

We propose a novel task called music score generation along with its benchmark system as follows.
As a result, we achieved a latent diffusion model capable of generating readable music score images conform to input text description.

\subsection{Music score generation system}
\label{sec:exp-setup}

The system founds on a text-driven latent diffusion model \cite{Rombach_2022_CVPR} as a generative engine which consists of a variational autoencoder \cite{kingma2013auto}, a text encoder and a UNet \cite{ronneberger2015unet} backbone.

\paragraph{VAE} We apply variational autoencoder (VAE) \cite{kingma2013auto} to encode the training images. The encoder $E$ encodes $x \in \mathbb{R}^{H \times W \times 3}$ in RGB space, into a latent representation $z = E(x)$. The latent representation $z$ has a shape of $[C, \frac{H}{f}, \frac{W}{f}]$, where $f$ represents the downsampling factor. In \cite{Rombach_2022_CVPR}, the suggested $f \in [4,16]$ yields the best results. Therefore, we chose the same value of $f = 8$ as the Stable Diffusion \cite{Rombach_2022_CVPR} official implementation.

\paragraph{Text encoder} In order to achieve our objective of guided generation using text, we utilize a text encoder to encode the text into embeddings that can serve as input to the model. The text encoder we use is OpenCLIP \cite{ilharco_gabriel_2021_5143773} in Stable Diffusion 2.0 \cite{Rombach_2022_CVPR} which an open source implementation of CLIP \cite{Radford2021LearningTV} by OpenAI. CLIP learns a multi-modal embedding space by jointly training an image encoder and text encoder to maximize the cosine similarity of the image and text embeddings \cite{Radford2021LearningTV}.
These embeddings serve as conditions to drive the UNet generation backbone.

\paragraph{Text-driven latent diffusion model} 

After calculating the text and image embeddings, UNet can be trained \cite{ronneberger2015unet, Rombach_2022_CVPR} as the generation engine in this system. We employ OpenCLIP \cite{ilharco_gabriel_2021_5143773, Radford2021LearningTV} to transform text into embeddings. The text-derived embeddings are fed into the UNet via spatial cross-attention mechanism \cite{Rombach_2022_CVPR, vaswani2017attention} for denoising generation process, yielding the latent representation of the corresponding music score images. This latent representation is then input into VAE decoder to obtain the final score image. This constitutes the complete text-to-score image process, playable score images can be generated through this process.

\paragraph{Training}

We fine-tuned a Stable Diffusion model \cite{Rombach_2022_CVPR} based on the stable-diffusion-2-base pre-trained weight (\texttt{512-base-ema.ckpt},\footnote{\url{https://huggingface.co/stabilityai/stable-diffusion-2-base}} SD2.0-base for short) by Stability AI. SD2.0-base is trained from scratch 550k steps at resolution $256 \times 256$ on a subset of LAION-5B \cite{schuhmann2022laion5b}, and is further trained for 850k steps at resolution $512 \times 512$ on the same dataset on images with a resolution larger than $ 512 \times 512$. The training consists of two phases. For the first phase, the VAE is fine-tuned using a dataset of 240 sheet piano and violin score images at a resolution of $512 \times 512$. We optimize VAE's parameters using Adam \cite{kingma2017adam} with a learning rate of $1 \times 10^{-5}$, a linear learning rate scheduler, no weight decay, and a batch size of 8. 

The second phase involves fine-tuning the UNet-based diffusion model. We utilize the MusicScore-400 subset to train the diffusion model while keeping the weights of the fine-tuned VAE frozen. The text encoder used is OpenCLIP-ViT/H \cite{ilharco_gabriel_2021_5143773, cherti2023reproducible, Radford2021LearningTV}. We select five categories from the metadata, including composer, instrumentation, piece style, key, and genre, to form the text condition. An example text condition is \textit{a music score, composer is [composer], instrumentation is [instrumentation], key is [key], piece style is [piece\_style], genre is [genre]}, where the values in the square brackets are arguments configured by users.
By training on MusicScore-400 subset at $ 512 \times 512$ resolution for 78,000 iterations, we achieve a music score generation system which is capable of generating playable music score image via text input. We conduct training using 8 RTX 4090 GPUs with a global batch size of 64. The gradient optimizer is AdamW \cite{loshchilov2019decoupled} without parameters modifications from PyTorch's \cite{NEURIPS2019_9015} implementation. The learning rate is $1 \times 10^{-5}$, no learning rate scheduler is applied.

\subsection{Evaluation}
\label{sec:exp-eval}

We measure the performance of music score generation using Fréchet Inception Distance (FID) \cite{heusel2018gans}, which is a standard metric for evaluating generative models of images. In Table~\ref{tab:vmg_fid}, FID-\textit{n} indicates we randomly select $n$ images for each evaluation, where $n \in [8, 16, 32, 64]$. The SD2.0-base is fine-tuned on $ 512 \times 512 $ resolution MusicScore-400 for 78,000 iterations. We evaluate FID \cite{Seitzer2020FID} across MusicScore-400, MusicScore-14k, and MusicScore-200k datasets, the ground-truth images from each subsets are resized to same resolution with generated images at $ 512 \times 512 $. During generation, we apply a DDIM sampler \cite{song2022denoising} for 250 DDIM sampling steps. The text prompt formats remain consistent between training and generation phases. We apply \textit{classifier-free guidance} \cite{ho2022classifier} with guidance strength $\omega = 4.0$ in generation. Table~\ref{tab:vmg_fid} presents the music generation results. We achieve FID scores of 74.46, 229.16, and 261.28 when evaluated on 64 images from the MusicScore-400, MusicScore-14k, and MusicScore-200k datasets, respectively. Figure~\ref{fig:txt2img_result} shows the generated music scores conditioned on text prompts. The first column displays a correctly generated A-major violin score. The third and fourth columns show correctly generated A-major piano scores, indicated by the number and pattern of sharp symbols \musSharp. 
\begin{figure*}[t!]
    \centering
    \begin{subfigure}[t]{0.23\textwidth}
        \centering
        \includegraphics[width=\textwidth]{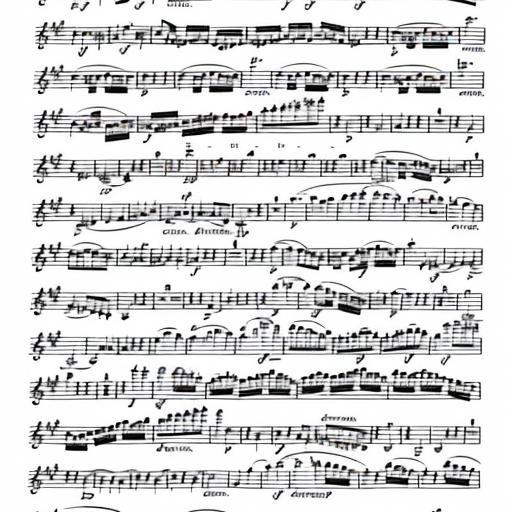}
        \caption{a music score, instrumentation is \textcolor{red}{violin}, key is \textcolor{blue}{A major}}
    \end{subfigure}%
    \hfill
    \begin{subfigure}[t]{0.23\textwidth}
        \centering
        \includegraphics[width=\textwidth]{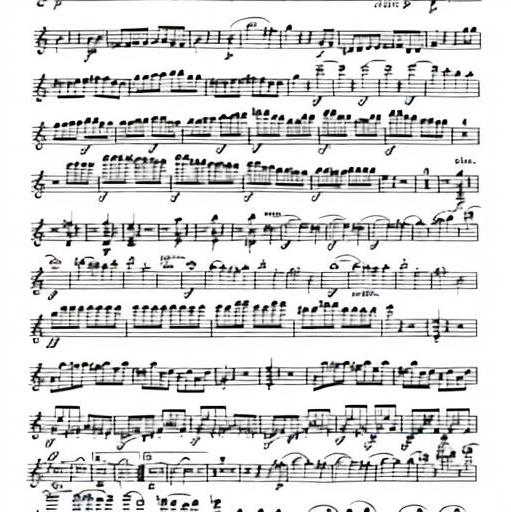}
        \caption{a music score, instrumentation is \textcolor{red}{violin}}
    \end{subfigure}
    \hfill
    \centering
    \begin{subfigure}[t]{0.23\textwidth}
        \centering
        \includegraphics[width=\textwidth]{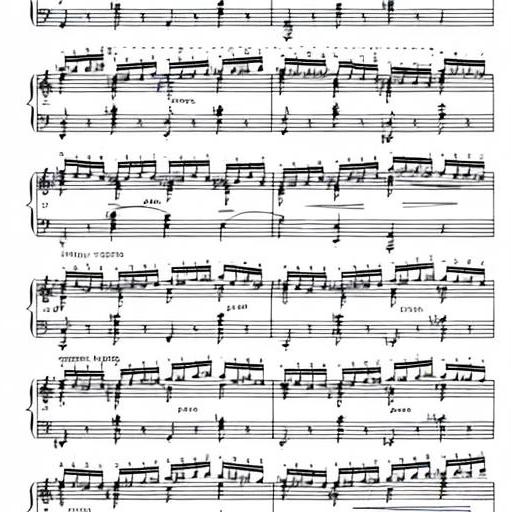}
        \caption{a music score, instrumentation is \textcolor{red}{piano}, key is \textcolor{blue}{A major}}
    \end{subfigure}%
    \hfill
    \begin{subfigure}[t]{0.23\textwidth}
        \centering
        \includegraphics[width=\textwidth]{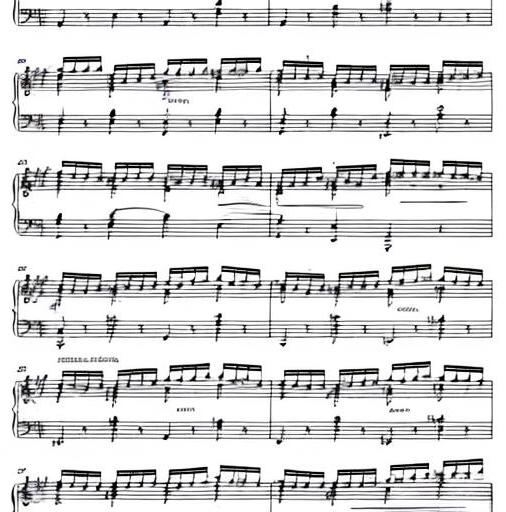}
        \caption{a music score, instrumentation is \textcolor{red}{piano}}
    \end{subfigure}
    \caption{Generated music scores. The selected samples are generated by the text-driven Stable Diffusion model fine-tuned on MusicScore-400 at $512 \times 512$. For all results, the \textit{classifier-free guidance} scale remains 4.0 and 250 DDIM sampling steps. The input text prompt follows the pattern of: \textit{a music score, instrumentation is [\textcolor{red}{instrumentation}], key is [\textcolor{blue}{key}]}.}
    \label{fig:txt2img_result}
\end{figure*}
\begin{table}[h]
  \caption{\textbf{Benchmarking text-conditional Stable Diffusion trained on MusicScore-400 with resolution $512 \times 512$.} FID-\textit{n} represents the number of evaluated images from respective subset of MusicScore.}
  \label{tab:vmg_fid}
  \centering
  \begin{tabular}{l|ccc}
    \toprule
    Subset& MusicScore-400 & MusicScore-14k & MusicScore-200k \\
    \midrule
    FID-8  & 114.65& 297.60 & 294.76 \\
    FID-16 & 85.81 & 221.42 & 314.06 \\
    FID-32 & 84.33 & 255.00 & 264.02 \\
    FID-64 & 74.46 & 229.16 & 261.28 \\
    \bottomrule
  \end{tabular}
\end{table}

\section{Limitations}
\label{sec:limit}

The MusicScore dataset consists of 32,307 PDF files, which were selected from a pool of 148,257 PDF files to ensure quality, resulting in a selection rate of 21.8\%. Consequently, some music scores from the original IMSLP pages may not be included in MusicScore. Second, MusicScore divides PDF files into individual JPG pages. The metadata describes the entire PDF file rather than individual JPG pages, potentially leading to mismatches between text and images. Third, the MusicScore dataset predominantly includes works by European composers, potentially creating an imbalance in the distribution of music works within the dataset.

\section{Conclusion}
\label{sec:conclude}

In this work, we introduce MusicScore, a large-scale music score dataset collected and processed from the International Music Score Library Project (IMSLP). MusicScore consists of image-text pairs, where the image represents a page of music score and the text contains metadata about the music. We provide details about the dataset collection and cleaning processes. MusicScore includes metadata extracted from the general information section of IMSLP pages, which covers composer, instrumentation, era, and other general information of the composition. MusicScore is curated into small, medium, and large scales, with 400, 14k, and 200k image-text pairs, respectively, offering various levels of diversity. MusicScore is publicly released. Additionally, we develop a music score generation system based on UNet diffusion models to generate music scores conditioned on text descriptions, aiming to benchmark MusicScore for music score generation tasks.
In the future, we plan to develop a MusicScore-CLIP model for music score modeling. We also intend to integrate score, audio, and symbolic representations to create unified music modeling and generation systems.

\newpage
\bibliographystyle{unsrt}
\bibliography{ref}

\newpage

\appendix

\section{MusicScore dataset}

\subsection{Dataset documentation}

MusicScore dataset is an image-text dataset. For the image and text of every single sample, the image is the music score page of classical music, the text is the metadata of the corresponding composition which this page belongs to. MusicScore dataset varies in three different scales, 403, 14656, and 204800 pairs of image and text data, namely, MusicScore-400, MusicScore-14k and MusicScore-200k respectively.

As MusicScore is an image-text pair dataset, the motivation of crafting MusicScore dataset is to encourage the connection between visual musical semantics and text description, which currently is still an unexplored territory. By exploring the potentials on the visual musical semantic information, equipping with appropriate usage of it may enhance music generation, even music understanding field in some degree.

MusicScore has been uploaded to Hugging Face: \url{https://huggingface.co/datasets/ZheqiDAI/MusicScore}. Three subsets in different scales are maintained, along with the Croissant metadata. For the scripts of color depth filtering and the PyTorch model code of score classification has also been uploaded to GitHub: \url{https://github.com/dzq84/MusicScore-script}.

\subsection{Intended use}

An example sample refers to Figure~\ref{fig:metadata}. The image is a single page of classical music score. The text can be organised by different attributes of metadata which recorded as a JSON file in terms of different scenarios and usages. Instead of providing processed sentence or paragraph, the reason of providing raw metadata as the text part is to maximize the ways of usage. The intended tasks where MusicScore dataset can be applied are text-to-image score generation, contrastive learning for music score images and text description. For music understanding, MusicScore can be used in a classification task, contrastive learning task between score image and text description.

\section{Experimental}

\subsection{Experiment setup}

\paragraph{Training data} There are two phases of fine-tuning in the experiment, fine-tuning VAE and UNet respectively. For fine-tuning VAE, the crafted dataset contains 240 images consist of 120 images of \textit{Preludes and Fugues through all tones and semitones} by Johann Sebastian Bach which also occurs in MusicScore-400 subset that used in fine-tuning UNet, and another 120 violin score images which are randomly selected from MusicScore-14k subset. For fine-tuning UNet, the dataset is MusicScore-400 subset that contains 403 images.

\paragraph{Hyperparameters} The hyperparameters used in the experiment is described in Table~\ref{tab:hyperparam}

\begin{table}[ht]
    \centering
    \begin{tabular}{c|c}
        \toprule
        \textit{UNet Configuration} & \\
        \midrule
        \textbf{Latent Dimension} $C_z$ & 4 \\
        \textbf{Downsample Factor} $f$ & 8 \\
        \textbf{Latent Scaling Factor} & 0.18215 \\
        \midrule
        \textit{Hyperparameters} & \\
        \midrule
        \textbf{Learning Rate} & $0.00001$ \\
        \textbf{Batch Size} & 8 \\
        \textbf{Adam $\epsilon$} & 1e-8 \\
        \textbf{Adam($\beta_1$, $\beta_2$)} & (0.9, 0.999) \\
        \textbf{Adam $\lambda$} &  0.01 \\
        \bottomrule
    \end{tabular}
    \caption{Detailed experimental settings.}
    \label{tab:hyperparam}
\end{table}

\paragraph{Distributed training} We use Distributed Data Parallel training throughout the whole experiment. The distribution is conducted by using Hugging Face's Accelerate, a library that simplifies distributed training of deep learning models, providing easy-to-use abstractions and utilities for efficient utilization of hardware resources. We use default Accelerate configuration which actually is a Data Distributed Parallel method. The training is conducted in FP32, without mixed precision training.

\begin{figure}[ht]
  \centering
  \includegraphics[width=1.0\textwidth]{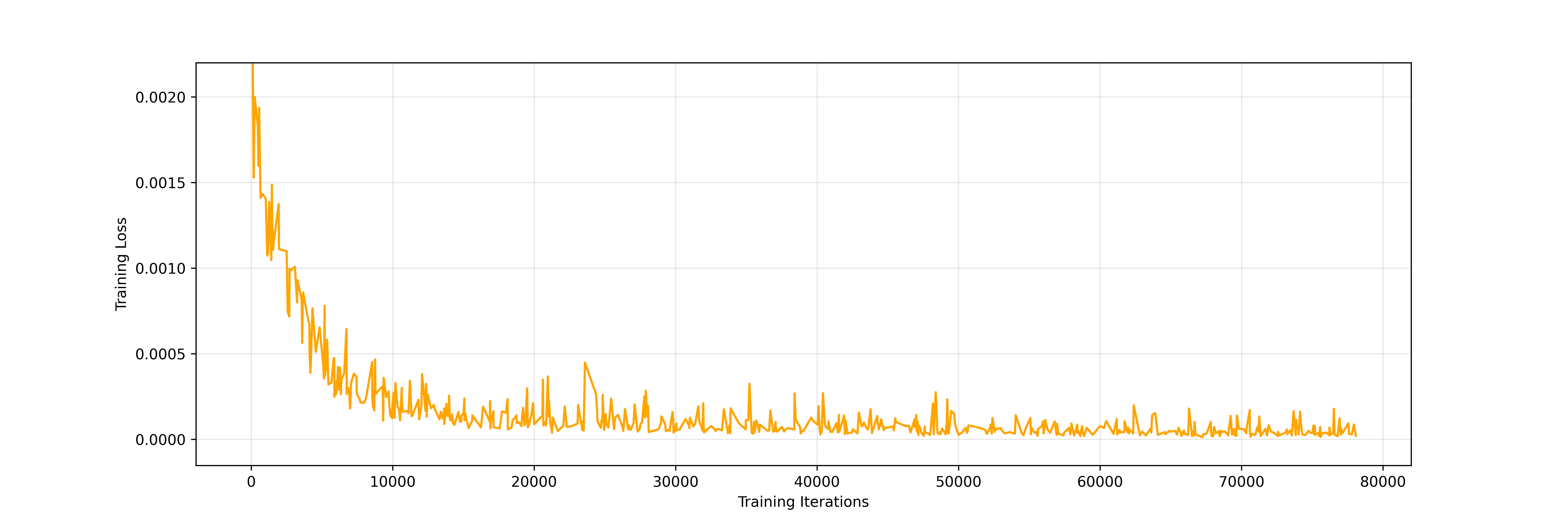}
  \caption{
  Training loss curve for fine-tuning UNet.
  }
  \label{fig:loss}
\end{figure}

We use single machine with 8 RTX 4090 GPUs both VAE and UNet fine-tuning. The loss curve for fine-tuning UNet refer to Figure~\ref{fig:loss}.

For diffusion, three prediction modes are optional for Stable Diffusion 2.0, which are $x_0$-prediction, $\epsilon$-prediction, and $v$-prediction. The $v$-prediction mode fuses prediction of $x_0$ and $\epsilon$ which considered as the state-of-the-art prediction method in many applications. In our experiment, we use $\epsilon$-prediction mode, i.e., to predict noise.

\subsection{Experiment result}

\begin{figure}[ht]
  \centering
  \includegraphics[width=1.0\textwidth]{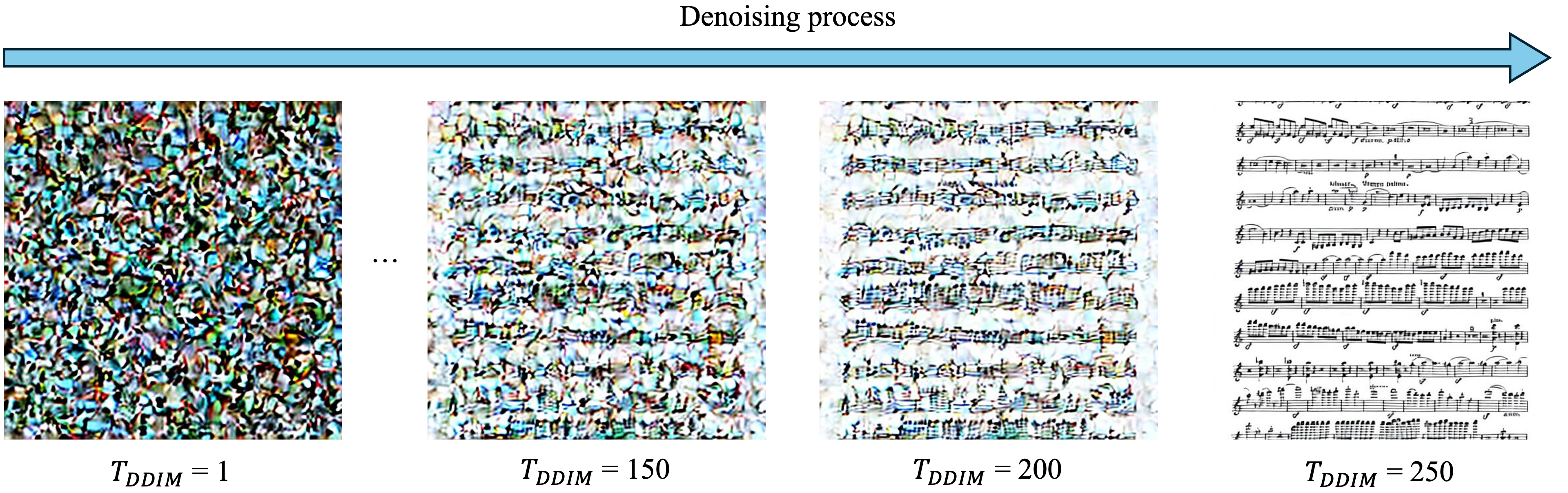}
  \caption{
  \textbf{Example of denoised intermediate images in denoising process.} The denoising sampler is DDIM Sampler with 250 DDIM sampling steps. The intermediates are saved every 50 steps. All results are guided using \textit{classifier-free guidance} of 4.0 \textit{cfg} scale.
  }
  \label{fig:denoise}
\end{figure}

Figure~\ref{fig:denoise} illustrates the denoising process of text-generated score image. We use DDIM sampler with a DDIM denoising steps of 250 to perform denoising process.

Figure~\ref{fig:txt2img_result} illustrates the text-generated score images at $512 \times 512$ resolution.

\clearpage

\begin{figure*}[t!]
    \centering
    \begin{subfigure}[t]{0.5\textwidth}
        \centering
        \includegraphics[width=\textwidth]{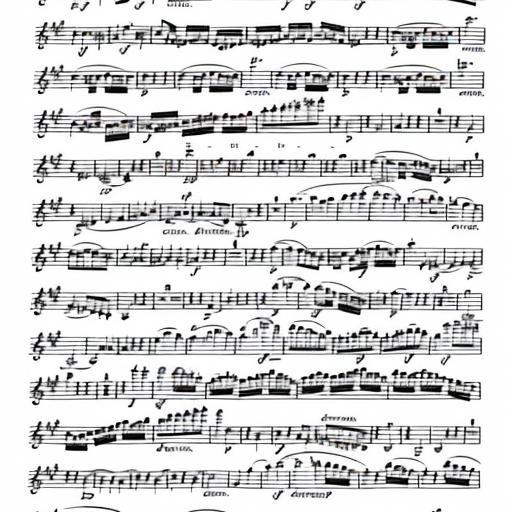}
        \caption{a music score, instrumentation is \textcolor{red}{violin}, key is \textcolor{blue}{A major}}
    \end{subfigure}%
    \hfill
    \begin{subfigure}[t]{0.5\textwidth}
        \centering
        \includegraphics[width=\textwidth]{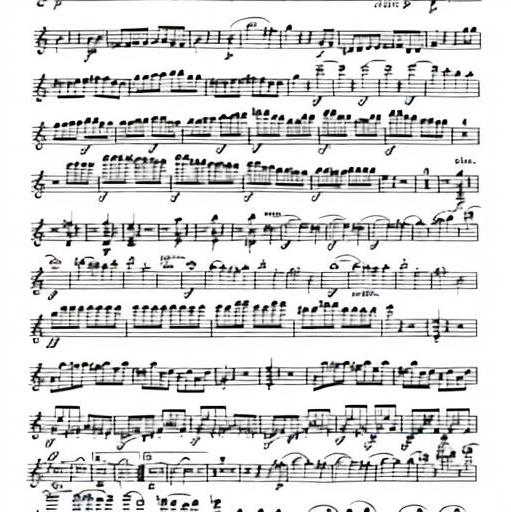}
        \caption{a music score, instrumentation is \textcolor{red}{violin}}
    \end{subfigure}
    \hfill
    \centering
    \begin{subfigure}[t]{0.5\textwidth}
        \centering
        \includegraphics[width=\textwidth]{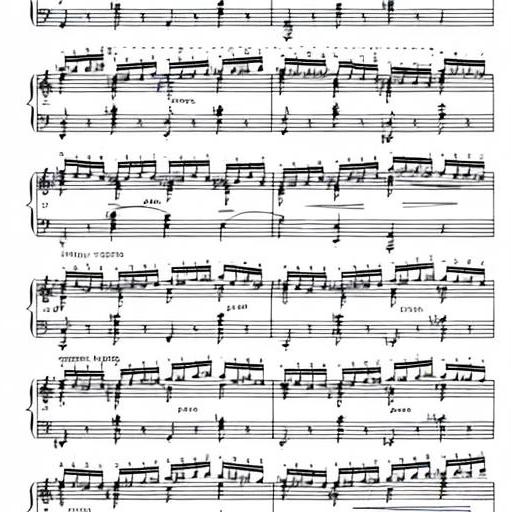}
        \caption{a music score, instrumentation is \textcolor{red}{piano}, key is \textcolor{blue}{A major}}
    \end{subfigure}%
    \hfill
    \begin{subfigure}[t]{0.5\textwidth}
        \centering
        \includegraphics[width=\textwidth]{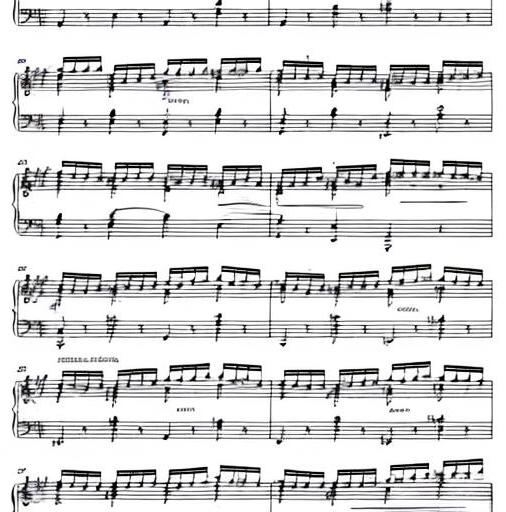}
        \caption{a music score, instrumentation is \textcolor{red}{piano}}
    \end{subfigure}
    \caption{Generated music scores. The selected samples are generated by the text-driven Stable Diffusion model fine-tuned on MusicScore-400 at $512 \times 512$. For all results, the \textit{classifier-free guidance} scale remains 4.0 and 250 DDIM sampling steps. The input text prompt follows the pattern of: \textit{a music score, instrumentation is [\textcolor{red}{instrumentation}], key is [\textcolor{blue}{key}]}.}
    \label{fig:txt2img_result}
\end{figure*}

\end{document}